\documentclass[aps,pre,reprint,groupedaddress]{revtex4-2}
\usepackage[tbtags]{mathtools}
\usepackage{amsmath,amsfonts,mathdots,amssymb,yfonts,calc}
\usepackage{graphicx}
\usepackage{subfigure}
\begin{document}
\title{Cyclically Symmetric Thomas Oscillators As Swarmalators : A paradigm for Active  Fluids \& Pattern Formation}
%*******************************************************
\author{Vinesh Vijayan}
\email[]{vinesh.phy@gmail.com}
\altaffiliation{N/A} %Department of Physics and Astronomy }
\affiliation{N/A}%National Institute of technology, Rourkela, India-769008}
%&&&&&&&&&&&&&&&&&&&&&&&&&&&&&&&&&&&&&777
\author{Pranaya Pratik Das}
\email[]{519ph1005@nitrkl.ac.in}
\altaffiliation{Department of Physics and Astronomy }
\affiliation{National Institute of technology, Rourkela, India-769008}
%$$$$$$$$$$$$$$$$$$$$$$$$$$$$$$$$$$$$$$

\date{\today}

\begin{abstract}
In this letter, we demonstrate the cyclically symmetric Thomas oscillators as swarmalators and describe their possible collective dynamics. We achieve this by sewing Kuromoto-type phase dynamics to particle dynamics represented by the Thomas model. More precisely, this is equivalent to a nonlinear particle aggregation model with cyclic symmetry of coordinates and position-dependent phase dynamics. The nonlinear equations describe spatiotemporal patterns of complex nature(irregular hexagonal lattices) and chaotic randomness(chaotic turbulence) at two extreme system parameter values. This pattern is the outcome of nonlinear self-organization, which leads to a new class of turbulent flow - active turbulence. We claim that this model can capture the dynamics of many naturally occurring microorganisms and microswimmers. The model described in this letter can be a prototypical model for understanding active systems and may shed light on the possibility of making novel materials(active matter) with exciting biomedical and industrial applications. The key to this is the understanding and control over the complex dynamics of active systems, an out-of-equilibrium system, which is potentially helpful in making functional materials, nano and micromachines.
\end{abstract}

%\maketitle must follow title, authors, abstract, and keyword
\maketitle
\tableofcontents

\section{\label{sc1}INTRODUCTION}
The recent attraction of research, which evoke much curiosity among researchers working on synchronization, is swarmalators. This phenomenon refers to many agents performing collective motion to achieve a specific process, like in the flocking and schooling behaviour of birds and fishes, insect aggregations, and migrations of cells. One can model them as oscillators that move collectively through space when coupled as opposed to laser arrays or cardiac pacemaker cells\cite{Kevin_p}\cite{Kev1}. These are oscillators whose phases affect their dynamics. As verified by experiments, the number of nearby oscillators and bidirectional coupling between spatial and phase dynamics controls the phase dynamics. These features play a crucial role in a population of bacteria whose internal dynamics are cyclically symmetric.\cite{Igo}. Like synchronization, swarming is also an emergent phenomenon, and achieving this new behaviour is known as self-organization. From a physics point of view, it opens up a wealth of new information and brings in a deep understanding of the general properties of active particles and self-propulsion. The mechanical properties of the particles in the medium and the statistical properties of the collective motion are equally crucial for a systematic understanding of living organisms and their motile counterparts, molecular motors. The collective motions observed in various situations point toward the interaction of complex oscillators to extremize specific processes trying to transport, dissipate and distribute energy efficiently.
\\
At the frontiers, the theoretical studies on the collective dynamics of swarmalators extended to that confined to moving on a ring in an attempt to understand bordertaxic vinegar eels and sperm\cite{Kev2}\cite{Kev3}. A nonlinear aggregation model is introduced in \cite{Yea19}, showing a positive minimal relative distance between particles and their uniform upper bound. Phase transitions occurring in a system of coupled swarmalators in 2D is the theme covered in\cite{Hon18}. The concept of a vision radius introduced in\cite{Kar22} helps to understand competitive time-varying phase interaction. The swarming behaviour is essential for understanding naturally occurring self-propelled particles and microswimmers. From an experimental point of view, it will also lead to the fabrication of novel materials, just like self-regulating soft robots\cite{Raj21}. Swarm robotics or micro-robotics is a new discipline where the ideas of swarming and cooperative dynamics are used for cargo delivery\cite{Akt22}\cite{Aga1}\cite{Aga2}. Since the particle size is small, one can use a swarm of such particles for the targeted delivery of drugs to cells. Soft micro-robots are particularly interesting because their trajectories can be controlled externally. Particularly interesting micro-robots are Xenobots which do not use any metal or semiconductors. Instead, they use biological tissues, and the advantage is that environmental pollution can be avoided compared to traditional micro-robots. They can consume energy from the surroundings and perform valuable work; these devices are biodegradable and bio-compatible\cite{Sam20}\cite{Phi20}\cite{Jos20}\cite{Dou21}. One controls all these cases by regulating the feedback mechanism\cite{Isl}.
\\
Rene Thomas proposed a straightforward 3D flow that represents, in the present context, a particle moving in a 3D lattice with frictional damping under the action of some source of momentum\cite{th099}. Due to its particle nature and local dynamics, they perform consecutive propagation and collision over a discrete lattice. One can think about the model as a particle with its immediate environment as a dynamical system. The model is given by
\begin{equation}
\begin{split}
\frac{dx}{dt} = -bx +siny\\
\frac{dy}{dt} = -by +sinz \\ 
\frac{dz}{dt} = -bz +sinx
\end{split}
\label{SE1}
\end{equation}
In a compact form one can write this as ${\bf v=f(r)}$ where, ${\bf r}=[x,y,z]^T$, ${\bf v}=[\dot{x},\dot{y},\dot{z}]^T$ and ${\bf f}=[f_1,f_2,f_3]^T$ and $\in \mathcal{R}^3$. The dot over the scripts shows the time derivative, and $f_1,f_2$ \& $ f_3$ are given in Equation(\ref{SE1}). Two essential characteristics of this system, the symmetry under the cyclic interchange of $x,y,z$ coordinates, and the other is sole parameter $b$, which is the frictional damping coefficient, need more attention. The former forms a feedback circuit regulating the particle's local dynamics, and the latter tunes the dynamics between a chaotic and regular motion. A low constant value of $b$ means a small number of lattice points, but when $b$ is high, the lattice is densely packed. Feedback loops make the Thomas system more attractive for understanding biological phenomena. These feedback mechanisms regulate the state variables involved. So it is worth mentioning to highlight the ability of Thomas's system to produce swarming behaviour, motility of micro-organisms and self-propulsion in general. For example, consider the dynamically interacting organisms and their environment. The production, cycling and regulation of energy and matter through the environment and the living organisms involve feedback circuits. The feedback circuit present in the Thomas system can be positive or negative depending on its state variables. The control parameter $b$ and the nature of feedback decide the system's 3D dynamics.
\\
Originally it was developed as a model for studying the role played by a feedback circuit in generating chaotic behaviour\cite{th099}. Theoretical models based on feedback circuits are useful for understanding the phenomena of many real systems like cell differentiation\cite{thomas-kauf2} and regulatory network\cite{thomas-kauf1}. It  represent many auto-catalytic chemical reactions\cite{Ram90}. In general, this system is suitable for the mathematical modelling of biological systems\cite{Vasi}\cite{vinesh1}. Sewing phase dynamics into a cyclically symmetric Thomas system will convert it into a swarmalator. The subsequnet dynamics and collective motion will be of great practical importance since it can model active fluids(dense suspension of bacteria, Artificial microswimmers).
\\
Consider a two-phase system with a dispersion medium and a dispersed phase(insoluble particles), and such a mixture is known as a colloid. As mentioned above, when the insoluble particles are swarmalators, we have an active colloidal system. Such colloidal suspensions are capable of executing functional tasks without external interventions\cite{Igor}\cite{Ebb}\cite{Ali}\cite{Jei}. To understand such behaviours, one needs to know the  environment and the interaction between the particles. The motion of active particles in a damping medium is exciting because it creates dynamic stresses in the medium and is the key to understanding much living matter\cite{Lis}\cite{Thom}. We assume that the size of the dispersed particles typically ranges from micrometre to nanometer range. The macroscopic properties of dispersions of extensive collections of such swarmalators are very well understood when the underlying interactions between individual particles are well understood\cite{Ben}\cite{Wei}\cite{Zou}.
\\
Particles that propagate and collide with discrete lattices at specific time intervals are a natural choice to understand multiphase/multicomponent complex flows. In the present context, we assume that the particle is moving in a 3D lattice under the action of an energy source. Then the little value of $b$ means that the spacing between the lattice points is significantly large. As a result, less frequent inelastic collision with the lattice points and shallow energy loss for the particle giving freedom to the particle to move around. For a high value of $b$, the 3D lattice will be densely packed, and the inelastic collision of the particle with the lattice will be frequent. As a result, it will lose energy and ultimately come to rest, making the motion impossible. When many such particles interact, they indicate the presence of a complex viscous flow. Due to movable and deformable interfaces, such flows are difficult to simulate with conventional computational techniques. These interfaces are manifestations of specific interactions among particles and with the environment. Depending on the medium parameters and particle size, the particle's motion may be under-damped or over-damped. When the Damping parameter is shallow inertial effects of the particle play a significant role, leading to Brownian activity. The inter-particle interactions are of great interest at the low damping of the medium since they decide the collective movement. This computational study demonstrates the interplay between particle interactions and medium parameters and portrays collective dynamics as a swarming process.  
\section{\label{sc2}Mathematical Modelling}
\subsection{The general framework and the swarming model}
This work considers a realistic version of swarmalators that are free to move in 3D space but in an environment with frictional damping. Swarming is achieved by sewing aggregation with synchronization. There is bidirectional coupling between phase and space dynamics and filling the necessary condition for realizing swarmalators. The dynamical equation, in general, is given by
\begin{equation}
\begin{aligned}
\frac{d{\bf r_i}}{dt}&={\bf v_i} +\frac{1}{N}\sum_{j\neq i}^N[{\bf I}_A({\bf r_j-r_i})\mathcal{F}_A(\Theta_j-\Theta_i)-{\bf I}_R({\bf r_j-r_i})\\
&\mathcal{F}_R(\Theta_j-\Theta_i) ]\\
\frac{d \Theta_i}{dt}&= \omega_i +\frac{K}{N}\sum_{j\neq i}^N \mathcal{H}(\Theta_j-\Theta_i)\mathcal{G}({\bf r_j-r_i})
\end{aligned}
\label{SE2}
\end{equation}
where ${\bf i}=1,....,N$ gives the population size. ${\bf r_i,v_i} \in \mathcal{R}^3$ and ${\bf r_i}=[x_i,y_i,z_i]^T$ is the position of the $i^{th}$ swarmalator at time $t$ and ${\bf v_i}$ gives its self propelsion velocity. ${\bf I}_A \hspace{0.2cm} \&  \hspace{0.2cm} {\bf I}_R$ captures the functional form of spatial interaction, attraction and repulsion, respectively, among swarmalators. $ \mathcal{F}_A$ and $\mathcal{F}_R$  stands for the functional influence of phase similarity on spatial attraction and repulsion. What makes it interesting is that the competition between their strengths lead to a conglomeration of particles with specific boundaries. $ \Theta_i$, $\omega_i$ gives the phase$(\Theta_i \in \mathcal{S}^1)$ and angular freqency of the $i^{th}$ swarmalator and $K$ gives the phase interaction strength. Physically phase means some internal property of the particles. There is spontaneous phase synchronization once a critical value of coupling constant $K_c$ is reached when uncoupled.
\\ 
For the specific 3D swarmalators, the following dynamical equations are used for simulation(Equation(\ref{SE3}). Cyclically symmetric Thomas system model({\bf f} is given by Equation(\ref{SE1})) for aggregation with attractive, repulsive coupling. That is, each particle possesses internal dynamics that vary cyclically. Kuramoto model is used to model synchronization\cite{Kur91}, but this time the phase dynamics is position dependent, giving the internal phase of the swarmalators where $\mathit{r} =|\bf{ r_j-r _i}|$, the distance between the $i^{th}$ and $j^{th}$ swarmalators.
\begin{equation}
\begin{aligned}
\frac{d{\bf r_i}}{dt}&={\bf f(r_i)} +\frac{1}{N}\sum_{j\neq i}^N[\frac{{\bf r_j-r_i}}{\mathit{r}}(A+JCOS(\Theta_j-\Theta_i))\\
& -B\frac{{\bf r_j-r_i}}{\mathit{r}^3}]\\
\frac{d \Theta_i}{dt}&= \frac{K}{N}\sum_{j\neq i}^N \frac{SIN(\Theta_j-\Theta_i)}{\mathit{r}^2}
\end{aligned}
\label{SE3}
\end{equation}
By comparing Equation(\ref{SE2}) and (\ref{SE3}) one can find out the functional nature of $ {\bf I}_A, {\bf I}_R ,\mathcal{F}_A, \mathcal{F}_R,\mathcal{H}$, and $\mathcal{G}$. From Equation(\ref{SE3}), one can observe two opposing forces act between two swarmalators: a short-range repulsive force and an attractive long-range force. When two of them are very near each other, the short-range repulsive force dominates and does not allow them to stick together or collide. On the other hand, the nature of attractive force is such that it depends on the phase difference between the swarmalators, controlled by the parameter J. One can see that the phase of swarmalators changes as a function of the phases of other swarmalators controlled by the parameter K. The magnitude of the phase-coupling force also varies with distance and hence depends on the positions of swarmalators. For this model, we assume that there is no role of phase similarity on spatial repulsion since $\mathcal{F}_R=1$. $\mathcal{F}_A$ is chosen to be $(A+Jcos(\Theta_j-\Theta_i))$ such that it is even and bounded.
\\
By rescaling the value of space and time, one can set the value of A and B to be one. When positive, the phase coupling strength K will tend to minimize the phase difference between swarmalators. And when it is negative, the phase difference increases.
We must keep $I_A$ always positive, constraining the J value between $\pm 1$. This parameter gives the extent to which phase synchrony increases spatial attraction. Now depending on the value of J, the following scenarios are observed. For $J>0$, swarmalators with the same phase will group in space; for $J<0$, the swarmalators with opposite phases will group. When J=0, their spatial attraction is independent of phase. For the Thomas system, we have one more parameter intrinsic to the system, $b$, which is the damping constant. By controlling this parameter, one can study the swarming process from a highly turbulent environment to a highly ordered fluid flow.
\\
\begin{table}
\centering
\caption{Different swarming states.}
\begin{tabular}{ |p{3.5cm}|p{2.3cm}|p{2cm}|  }
\hline
\multicolumn{3}{|c|}{ Static(1,2,3) \& Dynamic Swarming states(4,5).} \\
\hline
Asymptotic States & Parameter Values & Abbrevations \\
\hline
1.Static Sync &$J>0,K=1$& SS \\
2.Static Async &$J>0,K=-1$  & SA\\
3.Static Phase Wave &$J=1,K=0$& STPW \\
4.Splintered Phase wave & J=1,K$ \approx $0 & SPPW \\
5.Active Phase Wave &$J=1,K<0$ & APW\\
\hline
\end{tabular}
\label{T1}
\end{table}

\subsection{Characterization of the swarming states.}
Once the swarming states are obtained, one needs to distinguish between them using qualitative measures. Here we follow the order parameter formalism used in \cite{Kevin_p},\cite{Kar22}. The first step is to define the order parameter
\begin{equation}
\begin{aligned}
\mathcal{W}_{\pm} = \mathcal{S}_{\pm} e^{\Psi_{\pm}}
= \frac{1}{N}\sum_{j=1}^Ne^{j(\Phi_j \pm \Theta_j)}
\end{aligned}
\end{equation}
When there is a perfect correlation between the internal phases and spatial angle of the swarmalators, one can write $\phi_i=\pm \Theta_i + C$. The initial condition determines the $\pm$ and $C$. In this case, $s_{\pm} = 1$ and this value decreases when the correlation between $\Theta$ and $\phi$ is reduced. $s=Max(s_+,s_-)$ can be used to measure the correlation. So for $K=0$ we have $s_{\pm} = 1$(STPW), and for $K<0$, the correlation between  $\Theta$ and $\phi$  decreases. $s_{\pm}<1$ and varies non-monotonically with a further decrease in $K$. Once the APW is reached, the non-monotonicity disappears, and $s_{\pm}$ decreases uniformly until it finally reaches to zero.
\\
One can differentiate the stationary states from the non stationary states by measuring the mean velocity as follows for the 3D case.
\begin{equation}
\begin{aligned}
\mathcal{V} = \left\langle \frac{1}{N} \sum_{i=1}^N \sqrt{\dot{x}_i^2 +\dot{y}_i^2+\dot{z}_i^2+ \dot{\Theta}_i^2} \right\rangle_t
\end{aligned}
\end{equation}
Here the time average is taken. A finite non-zero value of the mean velocity indicates that the swarmalators are moving in space, and their phases vary in the interal $[0,2\pi)$. For stationary state $\mathcal{V}=0$. For different $J$ \& $K$, the swarmalators show different asymptotic states as mentioned in Table(\ref{T1}). Assuming all the mass at the origin initially, one can calculate the r.m.s displacement of the swarmalators using the formula
\begin{equation}
\begin{aligned}
\mathcal{D} =\left\langle  \sqrt{R^2} \right\rangle_t
\end{aligned}
\end{equation}
where $R=\frac{1}{N}\sum_{j=1}^N {\bf r}_j$ and a time average is taken. This will allow us to calculate the mobility of the swarming particles as a function of the damping parameter.
\section{\label{sc3}Numerical experiments and Observations}
In this work, we performed numerical experiments on swarmalators modelled by Equation(3). We run simulations by the RK4 method to solve the coupled differential equations. All swarmalators are initially in a cubical box with side two, and the phases of each of these oscillators were drawn uniformly from $[\pi,-\pi]$ at random. Then the subsequent collective dynamics are observed and analysed.
\\ 
At this point, the dynamics of an uncoupled Thomas oscillator are worth mentioning. The dynamics are controlled by the only parameter $b$, the damping parameter, indicating how much the system is dissipative. Depending on the value of this parameter, one can tune the dynamics from fixed points at higher values of $b$ to chaotic dynamics at the lower extreme($b$=0). The bifurcation diagram and Lyapunov spectrum are given below in FIG.\ref{F1}. For $b>1$, there is only one attractive fixed point for the system: the origin. The system has two attractive fixed points between $b = 1$ and $b = 0.328$. At $b=0.328$, the system undergoes a Hopf bifurcation, and a limit cycle is grown, and at $b=0.208$, the system becomes chaotic with quasi-periodic windows embedded in the chaotic sea.
\begin{figure}[hbt!]
    \centering
    \subfigure[]{\includegraphics[width=0.85\linewidth]{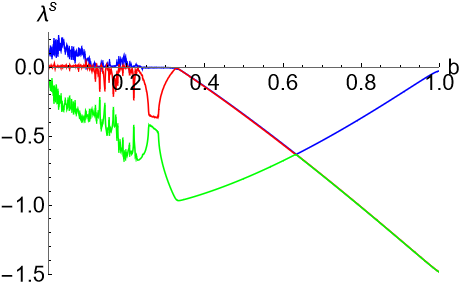}}     
    \caption{Lyapunov spectrum for Thomas oscillator.} 
 \label{F1}
\end{figure}

\subsection{Swarming States in the Dissipative Regime of Thomas Oscillator.}
\subsubsection{Swarming states for $b \in (0.328,1.2)$}
\begin{figure*}[hbt!]
    \centering
    \subfigure[DS]{\includegraphics[width=0.25\linewidth]{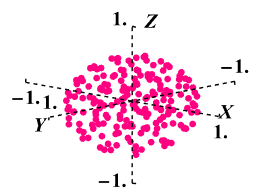}}
    \hfil
    \subfigure[SCAPW]{\includegraphics[width=0.25\linewidth]{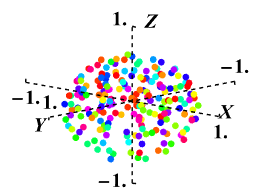}}\vfil    
    \subfigure[WCAPW]{\includegraphics[width=0.25\linewidth]{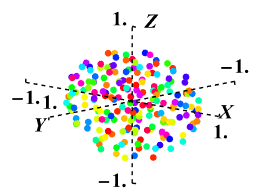}}
    \hfil 
    \subfigure[DAS]{\includegraphics[width=0.25\linewidth]{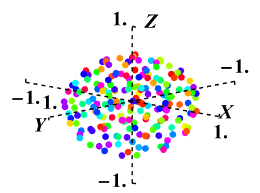}}\vfil
    \subfigure[DS]{\includegraphics[width=0.25\linewidth]{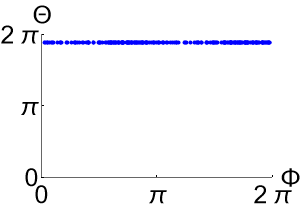}} 
    \hfil
    \subfigure[SCAPW]{\includegraphics[width=0.25\linewidth]{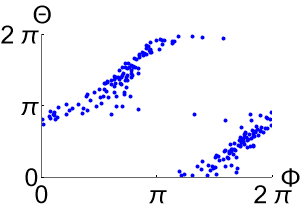}}\vfil       
    \subfigure[WCAPW]{\includegraphics[width=0.25\linewidth]{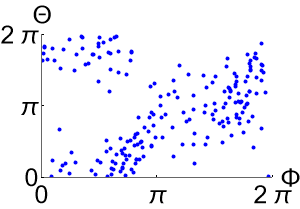}}
    \hfil 
    \subfigure[DAS]{\includegraphics[width=0.25\linewidth]{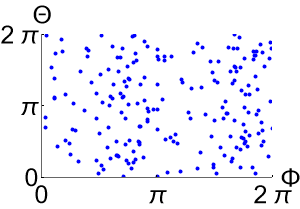}}  
\caption{ {\bf Top panel}: Swarming states in 3D Space. {\bf Bottom panel}: The $(\Phi, \Theta)$ distribution space for different swarming states. Simulations are done for $N=200$ for $T=1500 $ time units with $dT=0.01$ and $b=1.2$. (a,e)DS $(J,K)=(0.1,1)$,   (b,f)SCAPW $(J,K)=(1,0)$, (c,g)WCAPW $ (J,K)=(1,-0.3)$, (d,h) DAS  $(J,K)=(0.1,-1)$. } 
 \label{F2}
\end{figure*}

For $b>1$, i.e., for a high control parameter value, the uncoupled Thomas oscillator has only one equilibrium point, which is the attractive fixed points at ($x^*=y^*=z^*=0$   ). In this case, the collective motion of the swarmalators is such that they evolve into a circularly symmetric distribution in 3D space centred at the fixed point with an orientation. The viewpoint is so adjusted to show the circular symmetry explicitly. We observe the following scenarios for higher damping values $(b>1)$. First of all, there are no static swarming states. For example, when $J>1, K=1$, we observe Dynamic synchronization(DS)[FIG.\ref{F2}(a)] of swarmalators as opposed to the SS state reported in \cite{Kevin_p}. In the SS case, the swarmalators were fixed in space and had a single phase value asymptotically. For the case shown here, we observe the translational and rotational motion of the particles within the circular disc, and their phases switch asymptotically between two fixed values. One can observe a beautiful pattern within this circular disc - a manifestation of complex interactions, and it mimics an irregular hexagonal pattern. After the initial transients, at any time, the swarmalators are entirely synchronized in phase such that they adjust some of their internal property. These complex dynamics may be due to the continuous energy dissipation and gain happening locally due to the competition between phase-dependent space interaction and damping. The finite value of $\mathcal{V}$ verifies the dynamic nature. However, the circular disc as a whole is static.
\\
As the $K$ value decreases from zero, there is only an APW state. Since for $K=0$, there is a maximum correlation between phase and spatial variables, we call this swarming state to be Strongly Correlated APW(SCAPW), and for intermediate negative values of K, Weakly Correlated APW(WCAPW) as shown in FIG. \ref{F2}(b) and (c), respectively. When $K=0$, all the phases are frozen to their initial values. Since $J=1$, the positions of swarmalators rearrange such that like-phased swarmalators come nearer to each other. However, such a rearrangement is impossible due to strong frictional damping that competes with $J$. Instead, the swarmalators execute quivering motions about their mean position. It combines translation and clockwise/counter-clockwise rotation, resulting in an APW state. On further decreasing the value of $K$ from zero but still keeping it very near to zero, $K\approx0$, one can observe the amplitudes of oscillations become higher such that they start to oscillate in regular cycles both in space and phase. They also execute radial motion outward or inward along the disc. The system jumps to a Dynamic Async state(DAS) on further increase in the value of $K$ in the negative direction. In the DAS case, FIG. \ref{F2}(d), all the phases can occur and are shown with different colours of the oscillators. The swarmalators are dynamically$(\mathcal{V}\neq0)$ distributed uniformly over the disc, and every phase occurs everywhere. So as we decrease the value of $K$ from the positive to the negative side, the swarming pattern will be $DS \longrightarrow SCAPW \longrightarrow WCAPW \longrightarrow DAS$.\\
 
\begin{figure*}[hbt!]%height=4cm,width=4cm,angle = 0
    \centering
    \subfigure[DS]{\includegraphics[width=0.25\linewidth]{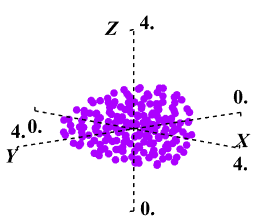}} 
    \hfil    
    \subfigure[SCAPW]{\includegraphics[width=0.25\linewidth]{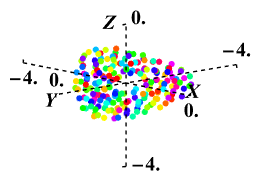}}\vfil
    \subfigure[MSPPW]{\includegraphics[width=0.25\linewidth]{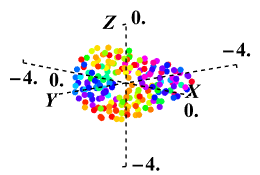}}
    \hfil    
    \subfigure[WCAPW]{\includegraphics[width=0.25\linewidth]{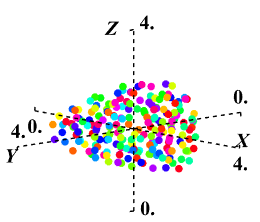}}\vfil
    \subfigure[DS]{\includegraphics[width=0.25\linewidth]{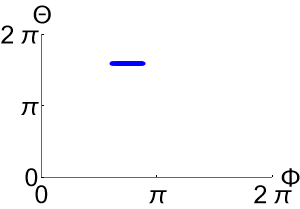}} 
    \hfil    
    \subfigure[SCAPW]{\includegraphics[width=0.25\linewidth]{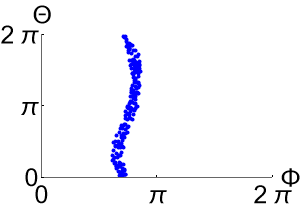}}\vfil
    \subfigure[MSPPW]{\includegraphics[width=0.25\linewidth]{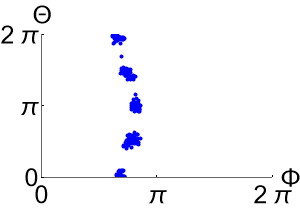}}
    \hfil    
    \subfigure[WCAPW]{\includegraphics[width=0.25\linewidth]{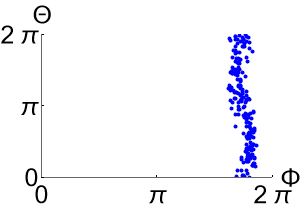}}  
\caption{{\bf Top panel}: Swarming states in 3D Space. {\bf Bottom panel}: The $(\Phi, \Theta)$ distribution space for different swarming states. Simulations are done for $N=200$ for $T=1500 $ time units with $dT=0.01$ and $b=0.5$. (a,e) DS $(J,K)=(0.1,1)$,  (b,f) SCAPW $(J,K)=(1,0)$, (c,g) MSPPW $(J,K)=(1,-0.03)$, (d,h) WCAPW $(J,K)=(1,-0.3)$.} 
 \label{F3}
\end{figure*}

\noindent The $(\Phi, \Theta)$ distributions are shown in the bottom panel of FIG.\ref{F2} for the swarming states depicted in the top panel. For DS, even though the swarmalators are moving within the circular disc, they all end up with the same phase $\Theta$ at a particular instant of time as in FIG.\ref{F2}(e), even if they all have different spatial orientations. FIG.\ref{F2}(h) shows the DAS distribution. Here the phase is uniformly distributed in all locations in space and shows uncorrelated behaviour of  $(\Phi, \Theta)$ distribution. FIG.\ref{F2}(f) and (g) show the correlation in the APW state. Here there is a strong correlation between phase and space angles, but the correlation decreases as the value $K$ increases in the negative direction.
\\
\noindent As mentioned earlier, we have one more extra parameter involved in the dynamics of the swarmalators. The first glimpse of its action manifested in the absence of static swarming states. For the particular case of $b>1$ we do not have STPW and SPPW for $(J=1, K=0)$ and $(J=1, K\approx 0^-)$, respectively. $b$, being a damping coefficient, opposes the attractive coupling strength $J$, which tends to bring the like-phased swarmalators together. The competition between these two strengths is such that there is a vigorous motion within the disc-shaped structure, indicating that the particles perform vigorous motion due to this competition(dynamic stress). So we have only an APW and SA on the negative side of $K$. Interestingly, such vigorous motions are observed in many biological swarms.
\\
\noindent At $b=1$, the uncoupled Thomas system undergoes a pitchfork bifurcation, resulting in two symmetrically placed attractive fixed points. The equilibrium points are given by $(x^*=y^*=z^*=\pm \sqrt{6(1-b)})$, and the collective dynamics are engaging in this case. The following scenarios are observed for the choice of damping coefficient $b=0.5$. The cluster formation of swarmalators is centred at the fixed points $\pm(1.73,1.73,1.73)$. They evolve into a Zindler curve-shaped disc[See APPENDIX] and to its mirror reflection about an axis parallel to the z-axis in 3D space with an orientation. The viewpoint is adjusted to demonstrate this symmetry explicitly. The choice among them may be energetically favourable. It is unclear what determines this choice, but it depends upon the interaction strengths and the number of oscillators. The observed scenario in this numerical experiment with two hundred swarmalators is such that when the swarms prefer the fixed point at the positive side, they evolve into a disc-shaped distribution with a mirror reflection of Zindler symmetry. In contrast, when they prefer the fixed point at the negative side, they own the Zindler symmetry as shown in FIG.\ref{F3}.
\\
We observe the DS  state with a complex dynamic pattern(irregular hexagonal lattices) for $K=1$ and $J>1$. When $K$ is reduced from zero to negative values, we have SCAPW( $K=0, J=1$) state followed by SPPW state. This SPPW differs from what is reported in \cite{Kevin_p}. The difference is that the separate clusters consist of phases near each other instead of a single phase. We call this swarming state a Mixed SPPW(MSPPW) to differentiate it from the conventional one. For higher negative values of $K$, we only have WCAPW. The role played by the damping coefficient is manifested for the second time here. Since the damping parameter is reduced by half, the competition between the damping force and phase-dependent position interaction strength is such that $J$ dominates. As we go from positive to negative values of $K$, we have $(DS \longrightarrow SCAPW \longrightarrow MSPPW \longrightarrow WCAPW)$. We tried to find a DAS state for $b=0.5$. FIG.\ref{F3} depicts the abovementioned scenarios with the corresponding $(\Theta,\Phi)$ distributions.\\

\begin{figure*}[hbt!]
    \centering
    \subfigure[DCS]{\includegraphics[width=0.35\linewidth]{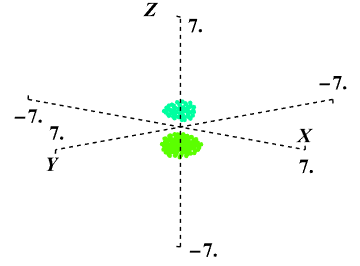}} 
    \hfil    
    \subfigure[$ (\Phi -  \Theta )$ Distribution]  {\includegraphics[width=0.35\linewidth]{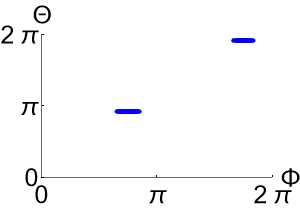}} 
      
\caption{(a) Dynamic Cluster Sync(DCS) and (b) the $(\Phi, \Theta)$ distribution space for the corresponding swarming state. Simulations are done for $N=200$ for $T=1500 $ time units with $dT=0.01$ and $b=0.5$. DCS $(J,K)=(1,\approx 0^+)$  } 
 \label{F4}
\end{figure*}

\noindent We also have the following observation from $(\Theta,\Phi)$ distribution for the swarming states. The spatial orientation of particles is confined to a small region along the $\Phi$-axis. For DS, all oscillators have the same phase value for a particular instant; for SCAPW, MSPPW, and WCAPW, the swarmalators move around with all possible phase values within the disc. The confinement indicates the particles' unique spatial orientation and directed motion within the cluster. Interestingly for $K=-1$ and $J>0$, the disc is located around $-1.73$, which may be an energetically favourable choice for the swarmalators, but the state is found to be WCAPW rather than DSA. A finite correlation between phase and spatial angles shows an APW state while entirely uncorrelated for DSA.
\\
\noindent The nature of swarming for a slightly positive value of $K$ leads to finding another swarming state, namely Dynamic Cluster Synchrony(DCS). Here one can see two distinct disc-shaped(Zindler cure-shaped) clusters with a mirror reflection symmetry. Within each cluster, there is complete phase synchrony. However, the clusters differ in phase asymptotically, and we say there is a phase asymmetry. Within each cluster, there is the quivering motion of particles which results in irregular hexagonal patterns. The number of swarmalators on each cluster depends sensitively on the value of $K$. The clusters and the corresponding phase-space correlations are shown in FIG.\ref{F4}.\\

\subsubsection{Swarming states for $b \in (0.0,0.328]$ } 
\begin{figure*}[hbt!]
    \centering
    \subfigure[t=25]{\includegraphics[width=0.3\linewidth]{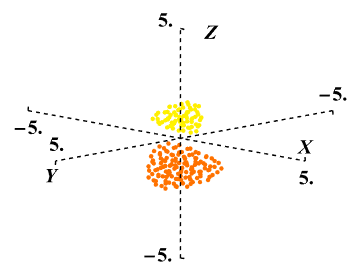}} 
    \hfil    
    \subfigure[t=140]{\includegraphics[width=0.3\linewidth]{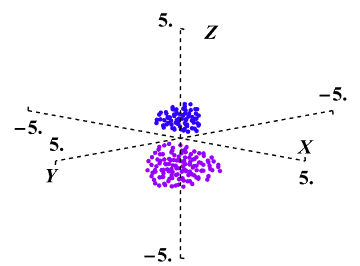}} \vfil
    \subfigure[t=240]{\includegraphics[width=0.3\linewidth]{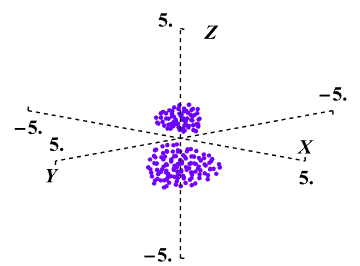}}
    \hfil
    \subfigure[t=430]{\includegraphics[width=0.3\linewidth]{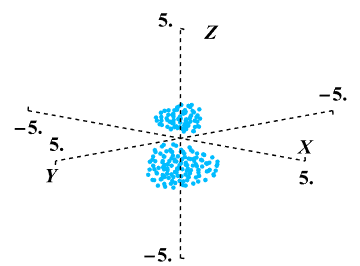}}     
    
    \caption{  Swarming in 3D Space for $b=0.3$ of Thomas oscillator for randomly selected time values.  Simulations are done for $N=200$ for $T=1500 $ time units with $dT=0.01$( $J=0.3$ and $K=1$).} 
 \label{F8}
\end{figure*}

\begin{figure}[hbt!]
    \centering       
    \subfigure[$ \Phi - \Theta $ \hspace{0.1cm} Distribution]  {\includegraphics[width=0.8\linewidth]{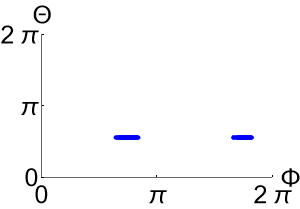}}     
\caption{(a) The $(\Phi, \Theta)$ distribution for the asymptotic swarming state shown in FIG.\ref{F8}(d). Simulations are done for $N=200$ for $T=1500 $ time units with $dT=0.01$ and $b=0.03$.} 
 \label{F11}
\end{figure}

This section considers the dynamically interesting domain of moderate damping parameter values. In particular, we choose $0.3$ and $0.1$. The reason for such a choice is that the former is within the limit cycle domain of the uncoupled Thomas oscillator, and the latter belongs to a well-defined chaotic regime[ref. FIG.\ref{F1}]. For further analysis, the parameter values are kept for complete phase synchronization( $J>0, K=1$). When $b=0.3$ and $J=0.3$, i.e., the two interaction strengths are balancing, the swarmalators split into two clusters and each cluster evolve into Zindler curve-shaped disc symmetry. The particles form complex patterns in each cluster(irregular hexagonal pattern) and remain asymptotically in DCS swarming state. The phase of clusters oscillates between two fixed values as shown in FIG.\ref{F8}. As we increase or decrease the $J$ value above or below this fixed value, particle migration exists from one cluster to the other. It asymptotically forms a single cluster in the DS swarming state. In comparison with the previous case of DCS, here we have phase symmetry between clusters as shown in FIG.\ref{F8} and \ref{F11}. This symmetry may be the result of the balancing of interaction strengths.\\

\begin{figure*}[hbt!]
    \centering
    \subfigure[t=45]{\includegraphics[width=0.3\linewidth]{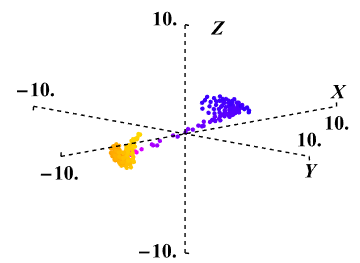}}
    \hfil     
    \subfigure[t=215]{\includegraphics[width=0.3\linewidth]{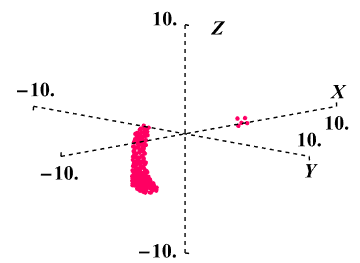}}\vfil
    \subfigure[t=325]{\includegraphics[width=0.3\linewidth]{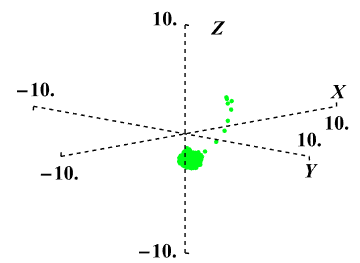}}
    \hfil
    \subfigure[t=590]{\includegraphics[width=0.3\linewidth]{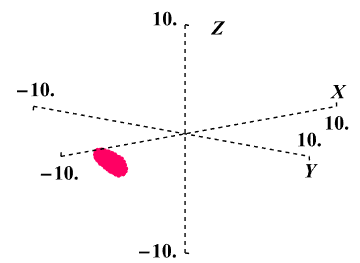}}     
    
    \caption{ Swarming in 3D Space for $b=0.1$ of Thomas oscillator for randomly selected time values.  Simulations are done for $N=200$ for $T=1500 $ time units with $dT=0.01$( $J=0.1$ and $K=1$). } 
 \label{F6}
\end{figure*}

\noindent For $b=0.1$, the particles, on the other hand, explore more volume than in the previous case. The collective motion is such that the particles split, recombine, and asymptotically reach a deformable disc shape: many times, only a few particles separate from the central disc but with the same phase. More interestingly, the collective motion is such that it behaves like a dense fluid. The scenario is such that as $b$ decreases, the particle velocity increases, and the particle cluster explores more space—the phase of the cluster oscillates between two fixed values and FIG. \ref{F6} depicts the swarming scenarios for randomly selected time values. An important point to note is that in the dynamic regime of Thomas oscillators, the swarming behaviour is such that there is a finite displacement of the cluster as opposed to the static regime where the cluster's centre of mass is stationary. The cluster as whole traces chaotic trajectory  in the 3D space.\\

\subsection{Swarming states for conservative  regime of Thomas oscillator [$b=0$]}
\begin{figure*}[hbt!]
    \centering
    \subfigure[t=30]{\includegraphics[width=0.3\linewidth]{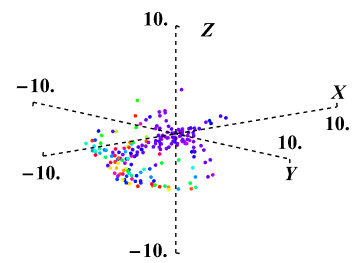}} 
    \hfil
    \subfigure[t=100]{\includegraphics[width=0.3\linewidth]{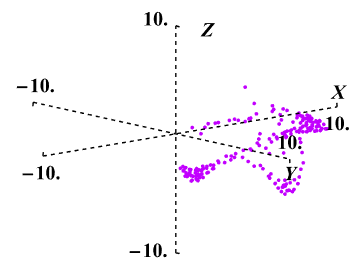}}\vfil
    \subfigure[t=150]{\includegraphics[width=0.3\linewidth]{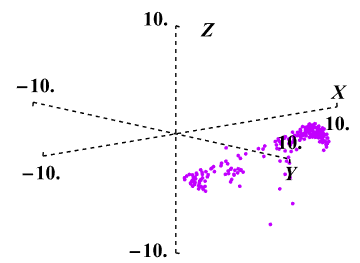}}
    \hfil     
    \subfigure[t=200]{\includegraphics[width=0.3\linewidth]{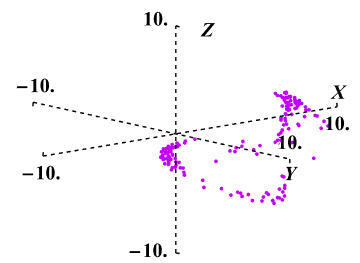}} 
    
\caption{  Swarming in 3D Space for the conservative regime of Thomas oscillator for randomly selected time values.  Simulations are done for $N=200$ for $T=1500 $ time units with $dT=0.01$ and $b=0$. We call the asymptotic swarming state as Dynamic Swarming Sync for $J=0.1$ and $K=1$. } 
 \label{F52}
\end{figure*}

\begin{figure*}[hbt!]
    \centering
    \subfigure[t=30]{\includegraphics[width=0.3\linewidth]{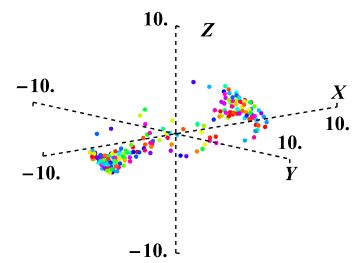}} 
    \hfil
    \subfigure[t=90]{\includegraphics[width=0.3\linewidth]{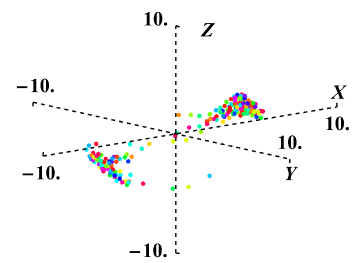}}\vfil
    \subfigure[t=140]{\includegraphics[width=0.3\linewidth]{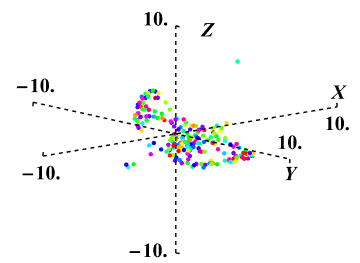}}
    \hfil     
    \subfigure[t=220]{\includegraphics[width=0.3\linewidth]{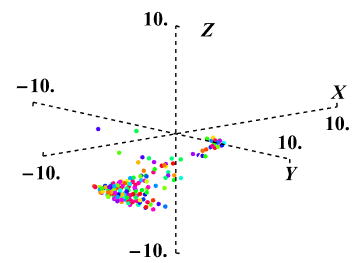}} 
    
\caption{  Swarming in 3D Space for the conservative regime of Thomas oscillator for randomly selected time values.  Simulations are done for $N=200$ for $T=1500 $ time units with $dT=0.01$ and $b=0$. We call the asymptotic swarming state as Dynamic Async for $J=0.1$ and $K=-1$. } 
 \label{F51}
\end{figure*}

Since the frictional damping term is zero, there is nothing to dissipate their motion; hence, the inertial forces dominate. The domination of inertial forces leads to collective action with large-scale swirling movement and exciting streaming patterns known as active turbulence. In the limit, $b=0$, the disclosed model effectively captures the physics of active turbulence. It is a special case of the Thomas system that for $b=0$, the system is conservative. In this case, the kinetic energy of the system is conserved. In this case, the swarmalator behaviour is exciting because they move around indefinitely in the 3D space. The particles' motions are such that they split and recombine by exploring more space than in the other case. The mean square displacement of the particles is much larger. When the $J$ value is high, the collective motion is such that the cohesive force between like-phased swarmalators is high, and they move around in a disc-shaped cluster. To distinguish this swarming behaviour from the previously noted ones,  we call these vigorous swarming states Dynamic Swarming Sync(DSS), when all the oscillators move around with the same phase, and Dynamic Active Swarming Sync(DASS) when they move around with different phases but with correlation. They both will have finite values of $\mathcal{V}$ and large values of $\mathcal{D}$. The range of parameter values are $(b=0,J>1,K=1)$ for DSS  and $(b=0,J>0,K=0,-ve)$ for DASS. These two scenarios are demonstrated in FIG.\ref{F51} and \ref{F52} \\

\begin{table*}
\centering
\caption{New swarming states.}
\begin{tabular}{ |p{5cm}|p{6cm}|p{3cm}|  }
\hline
\multicolumn{3}{|c|}{  Dynamic Swarming states} \\
\hline
Asymptotic States & Parameter Values & Abbrevations \\
\hline
1.Dynamic Sync &$b>0, J>0, K=1,\mathcal{V}\neq0,\mathcal{D}=0$& DS \\
2.Dynamic Async &$b>>0, J>0 ,K=-1,\mathcal{V}\neq0,\mathcal{D}=0$  & DAS\\
3.Dynamic Cluster Sync &$b>0, J=1, K=0^+,\mathcal{V}\neq0,\mathcal{D}=0$ &DCS\\
4.Dynamic Swarming Sync &$b \geqslant 0,J>1,K=+ve,\mathcal{V}\neq0,\mathcal{D}\neq0$& DSS\\
5.Dynamic Active Sarming Sync &$b\geqslant 0, J=1,K=0,-ve,\mathcal{V}\neq0,\mathcal{D}\neq0$&DASS\\ 
\hline
\end{tabular}
\label{T2}
\end{table*}

\section{\label{sc4}Results and Discussions}
As a swarmalator model, the Thomas system is an active fluid system. Here, the self-propelled particles can move around in a medium whose density depends on controlling the damping parameter. At high damping parameter values, the particles become trapped and remain static. They form a highly ordered(crystalline) disc-shaped structure in the fluid around the equilibrium point. The system's intrinsic dynamics and various parameters determine the collective mass's dynamics, phase and structure. At lower densities, i.e., low value of the damping parameter, the particles behave like a fluid and move around collectively within the medium. One can talk about active Brownian motion and associated transport phenomena during this non-equilibrium process. This also opens up the possibility of materials with unconventional properties.
\\
One can draw a close analogy between bacterial colonies and such active colloidal systems. For example, the formation of the SS state can be interpreted as the auto-aggregation of genetically similar species when they clump together. On the other hand, forming the SA state is a co-aggregation of bacteria where genetically distinct species clumps together. Microorganisms in a wide range of enterprises in the ecosystem are highly structured and complex, as observed in these cases. Most bacterial colonies exist in fluctuating environments, where shear forces, competition for nutrients, etc.,   may lead to the APW state's formation. This can be visualized as dynamic stress within the cluster resulting from the competition of various forces involved  and is a key to understanding living matter. One can think about the specific nature and strength of the interaction forces leading to the formation of these complex structures by natural selection. At the level of microorganisms, forming such intricate designs is a survival policy by staying together for food and saving from predators without the aid of a central organizer. From a physical point of view, nature is trying to minimize energy during its transport and dissipation.
\\
Forming biofilm is a dynamic process involving bacterial adhesion and clustering. The microorganisms organize into a coordinated functional entity and are found to be attached to living/non-living surfaces. They may contain single species or a diverse group of organisms(bacteria, archaea, protozoa, fungi and algae), each endowed with its metabolic function. In a natural environment, microorganisms thrive in complex communities. The survival of a single organism depends on the short-range interactions with other cells in the population and is responsible for the biofilm. The efficiency of a process in such a complex community depends on shared energy goods which allow them to perform a shared task. Microbial communities are efficiently and productively used in many biotechnological applications. Microbial processes are critical in ecological and evolutionary dynamics underlying their prevalence in natural, industrial and medical settings.
\\
A dynamically interacting system of organisms(communities they make up) and components of their environment form an ecosystem. Its primary process is the production of nutrients and their cycling and regulating matter and energy in its environment. this is a process where there is no energy loss due to dissipation. In the limit $b=0$, the swarmalators model the collective dynamics of such biological systems. Regarding the swimming of microorganisms, moving forward in a fluid via momentum conservation is ineffective. Because of their small sizes, the inertial effects will be subsided by the viscosity of the liquid. However still, they are equipped with efficient swimming mechanisms. Usually, the swimming of the active particles is due to low Raynolds number hydrodynamics and thermal noise. The physical structure of the particle can also trigger motion. Absorption energy from the surroundings creates field gradients around the particles inducing self-phoretic action. The model discussed in the paper does not involve any moving parts. Even then, they achieve self-propulsion and swimming by interacting with the environment, absorbing energy, and converting that into motion. In the model, the interaction with the environment is through the damping parameter. This interaction causes thermal fluctuations in the system, which in turn causes rotational and translational motion. The microswimmer moves persistently in all directions, collectively  covering a much larger area and hence has larger diffusivity. When the damping of the medium is zero, the inertial effects dominate. Since no buoyant forces are present, the particles will be in vigorous motion following the cyclic symmetry of the model. As observed in the simulations, the microswimmers swim together due to the exchange of information and behave cooperatively, followed by complex behaviour. During their collective motion, they re-organize via an internal feedback mechanism. They can perform movement along a straight line, move along a circular path, make sharp turns, and get back to the initial position.
\\
One final point to discuss is adaptation. Adapting to environmental changes plays a vital role in the survival of bacteria triggered by their genetics. From biology, in a more dense setting, the bacteria adapt by forming what is known as swarm cells. They adopt a different means of locomotion than individual bacteria. In the model discussed in the paper, a similar scenario follows. When parameter $b$ is tuned from zero to some finite value, the swarmalators form a disc-shaped film in 3D space and move collectively. Finally, when the damping is very high, they form a crystalline structure and remain static. Thus if the environment is exposed to continuous changes, swarmalators respond to it and adapt. Thus, the model helps us understand the fundamental physical mechanism governing self-assembly and collective behaviour in active matter systems and its natural tendency to organize into complex functional architectures. Examples range from tunable self-healing colloidal crystals and self-assembled microorganisms and robots. This disclosed model exhibit properties that their passive counterparts do not show.
\section{\label{sc5}Conclusion}
According to modern-day physics, active materials are considered free energy-consuming soft matter and give new directions in physics and biology. The collective motion happens spontaneously when the interaction forces are switched on in an otherwise static distribution of particles. The particles are coupled to an energy source, and the energy consumed is converted into work sustainably. Cyclically symmetric Thomas Oscillators as swarmalators are exciting options for understanding dynamics and pattern formation in an active fluid system. They show unusual macroscopic structures, including a transition from random swarming motion to a highly complex irregular hexagonal lattice structure. These behaviours are intimately connected to external parameters and the internal dynamics of individual oscillators. In this study, we could also find a few new swarming states. Naturally occurring biological machines are very efficiently designed by nature; probing and investigating them may lead to a better understanding and insight into non-equilibrium physics and point towards novel applications in physics, chemistry, biology, and engineering.
\section*{Acknowledgements}
Wish to acknowledge the support and motivation from Mr Ralu Johny and Mr Ratheesh S Chandran.
\section{\label{A1}Appendix}
\subsection{The Zindler Curve}
Consider the following parametric equation with a single real parameter $\epsilon$ for a simple mathematical illustration. 
\begin{equation}
\begin{aligned}
Z(w) &= X(w) + i Y(w)\\
     &= e^{i2w}+2e^{-iw}+\epsilon e^{iw/2}\hspace{1cm} w\in[0,4\pi]
\end{aligned}
\label{AE1}
\end{equation}
The Zindler curve is defined as a closed curve in a plane such that all chords which cut the curve length into equal halves have the same length. For $\epsilon > 4$, the curve is a Zindler Curve. In order to prove this the derivative of Equation(\ref{AE1}) and its absolute value are considered as given below.
\begin{equation}
\begin{aligned}
Z'(w) &= i(2e^{i2w}-2e^{-iw}+\frac{\epsilon}{2}e^{iw/2})\\
|Z'(w)|^2&= Z'(w)Z'(w)^*= 8+\frac{\epsilon^2}{2}-8cos3
\end{aligned}
\label{AE2}
\end{equation}
From Equation(\ref{AE2}) it is clear that $|Z'(w)|$ is $2\pi$-periodic. Then for any fixed value of the variable, say $w_0$, the following equation holds, which is half the length of the entire curve.
\begin{equation}
\begin{aligned}
\int_{w_0}^{w_0+2\pi} |Z'(w)|dw=\int_0^{2\pi}|Z'(w)|dw
\end{aligned}
\label{AE3}
\end{equation}
The length of the straight line, which divides the curve into two halves, can be shown to be independent of $w_0$, the reference point. These straight lines are bounded by the points $Z(w_0)$ and $Z(w_0+2\pi)$ for any choice of the reference point $w_0 \in[0,4\pi]$. The length of such a straight line is
\begin{equation}
\begin{aligned}
|Z(w_0+2\pi)- Z(w_0)|=|2\epsilon e^{iw_0/2}|=2\epsilon
\end{aligned}
\label{AE4}
\end{equation}
The following plots show the curves for a different choice of the parameter $ \epsilon$. For $\epsilon=4$, the chord meets the curve at one additional point and is not a Zindler curve. The simplest of all Zindler curves is a circle\cite{Jbr}\cite{Dav}.
\begin{figure*}[hbt!]
    \centering
\subfigure[]{\includegraphics[width=0.35\linewidth]{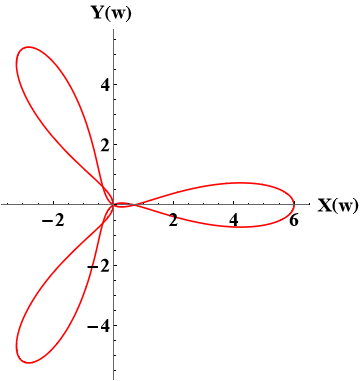}}
    \hfil 
\subfigure[]{\includegraphics[width=0.35\linewidth]{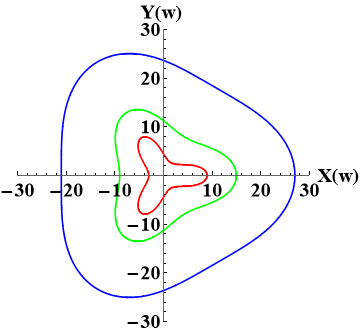}}    
    \caption{(a) Shows the curve for $\epsilon=3$ and is not a Zindler curve and (b) depicts the Zindler curves for $\epsilon=8, 16, 24$(red,green,blue) respectively.} 
 \label{FA}
\end{figure*}
Figure (\ref{FA}) shows the behaviour of Equation (\ref{AE1}) for four different values of the parameter $\epsilon$. 
\newpage
\bibliography{Bibliography}

\end{document}